\documentclass[aps,prl,twocolumn,showpacs]{revtex4}

\newcommand{\be}{\begin{equation}}
\newcommand{\ee}{\end{equation}}
\newcommand{\bea}{\begin{eqnarray}}
\newcommand{\eea}{\end{eqnarray}}
\newcommand{\bt}{\begin{tabular}}
\newcommand{\et}{\end{tabular}}
\newcommand{\ba}{\begin{array}}
\newcommand{\ea}{\end{array}}

\newcommand{\bvec}{\mathbf}

\begin{document}


$~$ \newline $~$
\hfill{}{{\small 
DSF$-$38/2004}}


\title{Quantum phase excitations in Ginzburg-Landau superconductors}

\author{E. Di Grezia}
\author{S. Esposito}
\email{Salvatore.Esposito@na.infn.it}
\affiliation{Dipartimento di Scienze Fisiche, Universit\`{a} di Napoli ``Federico II'' and Istituto Nazionale di Fisica Nucleare,
Sezione di Napoli, Complesso Universitario di Monte S. Angelo,
Via Cinthia, I-80126 Napoli, Italy}
\author{A. Naddeo}
\affiliation{Dipartimento di Scienze Fisiche, Universit\`{a} di Napoli ``Federico II'' and Coherentia-INFM, Unit\'a di Napoli, Complesso Universitario di Monte S. Angelo, Via Cinthia, I-80126 Napoli, Italy}

\begin{abstract}
We give a straightforward generalization of the
Ginzburg-Landau theory for superconductors where the scalar phase
field is replaced by an antisymmetric Kalb-Ramond field. 
We predict that at very low temperatures, where quantum phase effects are expected to play a significant role, the presence of vortices destroys superconductivity. 
\end{abstract}


\maketitle



The concept of order was introduced in condensed matter theory \cite{anderson} in the phenomenon of
spontaneous breaking of a symmetry \cite{nambu} characterized by a non-zero local order parameter (the vacuum expectation
value of a local operator). 
Spontaneous symmetry breaking takes place when an invariance of the Hamiltonian of a physical system is not respected by its
ground state. Symmetry is thus a universal property
featuring different phases of the system under study. According to the Goldstone theorem, for every
spontaneously broken global continuous symmetry, a
massless particle (the Goldstone boson) comes out \cite{nambu}; hence the spontaneous symmetry breakdown paradigm provides the origin of many gapless excitations, such as phonons and spin waves, which give rise to the low energy properties of many systems. In a gauge invariant theory with spontaneous symmetry breaking, after a particular gauge
transformation the Goldstone bosons disappear and the gauge bosons become massive.
This is the so called Anderson-Higgs mechanism
\cite {higgs} which plays a fundamental role in the theory of
gapped BCS superconductors: it is in fact the conceptual paradigm
underlying the effective Ginzburg-Landau (GL) theory
\cite{tinkham}.
\\
On 1982 the discovery of fractional quantum Hall (FQH) states
\cite{stormer} has brought to the attention of physicists a new
striking feature: such new systems contain many different phases
at zero temperature which display the same symmetry and then
cannot be classified through their symmetries. As a consequence,
long range correlations are not present and these systems cannot be described by the GL theory. A weak form of order (topological order) was proposed for FQH states \cite{wen} and then widely invoked in the discussion of a number of exotic phases exhibited by strongly correlated systems such as cuprate
superconductors \cite{fisher}, quantum dimer models
\cite{kivelson}, Josephson junction networks
\cite{sodano} and also three-dimensional ordinary superconductors \cite{balach}.
The concept of topological order applies only to states with finite energy gap and it has been recently generalized to quantum order \cite{wen1} for gapless quantum states. Now, topological field theory seems to provide a natural framework for topologically ordered states. Depending on the space-time
dimensions, Chern-Simons \cite{chern} or $BF-$terms \cite{blau}
will appear in the low energy effective action.

In this letter we give a generalization of the Anderson-Higgs mechanism
for ordinary BCS superconductors in $\left( 3+1\right)
$ space-time dimensions. The task will be pursued by means of the
well established equivalence \cite {lund} between the degrees of
freedom of a massless Kalb-Ramond (KR) field
\cite{kalb}, \cite{salvatore} and the ones of a (real) scalar massless field. The net result will be an effective lagrangian describing the system at energies well below the Higgs mass (i.e. at temperatures well below $T_{c}$), where superconductivity is washed out by a conserved topological current. Furthermore the gauge boson is gapless, which could be a signal of topological order \cite{wen1}.

In the GL theory \cite{tinkham} the Cooper pairs are
described by a (complex) scalar field $\phi $, whose lagrangian
density is the following:
\begin{equation}
{\cal L}=\left( \partial_{\mu }\phi \right)^{\dagger }\left(
\partial^{\mu}\phi \right) -U(\phi ,\phi^{\dagger }),
\end{equation}
where the potential energy is parameterized in terms of a coupling
constant $\lambda$ and a value $\phi_{0}$ corresponding to its
minimum: $U(\phi ,\phi^{\dagger })=\lambda (|\phi|^{2}-\phi_{0}^{2})$.
The dynamical EL equation for $\phi $,
considering only terms up to the quadratic ones, reads as:
\begin{equation}
\partial^{\mu }\partial_{\mu }\phi +m^{2}\phi \simeq 0, \label{4}
\end{equation}
with $m^{2}=-2\lambda \phi_{0}^{2}$, being the mass parameter.
In the simple one-dimensional case (along the $x$%
-axis), the solution of Eq. (\ref{4}) has the form:
\begin{equation}
\phi (x)\simeq \phi (0) \, e^{-x/\xi},  \label{4b}
\end{equation}
with a coherence length
\begin{equation}
\xi =\frac{1}{\sqrt{m^{2}}}=\frac{\xi_0 }{\sqrt{\frac{T}{T_{c}}-1}}.
\label{4c}
\end{equation}
This solution then describes the superconductor at temperature $T$
greater than the critical one: $T>T_{c}$. In this formalism, the
electromagnetic interactions of the Cooper pairs, mediated by a
gauge potential $A_{\mu }$, are obtained by means of the minimal
coupling prescription, replacing the derivative $\partial_{\mu }$
with the covariant derivative $D_{\mu }=\partial_{\mu }+2ieA_{\mu
}$.
The complete lagrangian of the system is then the following:
\begin{equation}
{\cal L}_{GL}=-\frac{1}{4}F_{\mu \nu }F^{\mu \nu }+\left( D_{\mu }\phi
\right)^{\dagger }\left( D^{\mu }\phi \right) -U(\phi ,\phi^{\dagger }), \label{7}
\end{equation}
where the first term ($F_{\mu \nu }=\partial_{\mu }A_{\nu
}- \partial_{\nu }A_{\mu }$ being the electromagnetic field
strength) accounts for the kinetic energy of the electromagnetic
field. The dynamical equations for the scalar and the
electromagnetic field become:
\begin{eqnarray}
&&
\partial^{\mu }\partial_{\mu }\phi
+m^{2}\phi +2\lambda |\phi |^{2}\phi =  \nonumber \\
&&
-2ie \left[(\partial^{\mu
}A_{\nu
})\phi + 2A^{\mu }\partial_{\mu }\phi \right] + 4e^{2}A^{\mu }A_{\mu }\phi , \\
&&\partial_{\mu }F^{\mu \nu }=-2ie(\phi \partial^{\nu }\phi^{\dagger
}-\phi^{\dagger }\partial^{\nu }\phi )-8e^{2}A^{\nu }|\phi |^{2}.
\label{8}
\end{eqnarray}
Let us now study the small fluctuations of the scalar field around its
minimum $\phi_{0}$ and expand $\phi $ as follows:
\begin{equation}
\phi (x)=\rho (x)e^{i\theta (x)}=(\phi_{0}+\eta (x))e^{i\theta (x)},
\label{10}
\end{equation}
where $\theta (x)$ is the (real) phase field, while the (real) $\eta (x)$
field properly describes the fluctuations of the system. By substituting Eq.
(\ref{10}) in the lagrangian (\ref{7}), and washing out an unphysical gauge
transformation\footnote{A gauge transformation does not affect the physics of the system as long as the field $A_{\mu }$ is massless; see below.} $A_{\mu}\rightarrow A_{\mu }+\frac{1}{2e}\partial_{\mu }\theta $, we have:
\be
{\cal L}_{GL}={\cal L}_{A}+{\cal L}_{\eta }+{\cal L}_{h.o.},
\label{11}
\ee
\begin{eqnarray}
{\cal L}_{A}&=&-\frac{1}{4}F_{\mu \nu }F^{\mu \nu }+4e^{2}\phi
_{0}^{2}A_{\mu }A^{\mu },  \label{12} \\
{\cal L}_{\eta }&=&(\partial_{\mu }\eta )(\partial^{\mu }\eta
)-4\lambda
\phi_{0}^{2}\eta^{2}, \\
{\cal L}_{h.o.}&=&-4\lambda \phi_{0}\eta^{3}-\lambda
\eta^{4}+8e^{2}\phi_{0}\eta A_{\mu }A^{\mu } \nonumber \\
&~& +4e^{2}\eta^{2}A_{\mu }A^{\mu }. \label{14}
\end{eqnarray}
The last part in ${\cal L}_{GL}$ accounts for higher order terms (third
and fourth order terms in the fields) and, for small fluctuations,
will be not considered here. The lagrangian ${\cal L}_{A}$ in
(\ref{12}) is that typical of a photon with mass $m_{A}$ given by
$m_{A}^{2}=8e^{2}\phi_{0}^{2}$, while ${\cal L}_\eta$ describes a
(real) scalar field $\eta$ with a mass $m_\eta$ given by $m_{\eta
}^{2}=4\lambda \phi_{0}^{2}=-2m^{2}$. Note that the phase field
$\theta $ has disappeared from our description \cite{higgs}, due
to the gauge transformation mentioned above. The physical picture
is, then, the following. For $T>T_{c}$ the condensate of the
Cooper pairs can be described by a complex scalar field $\phi $
with mass parameter $m^{2}$, while the infinite-range
electromagnetic interactions are described by a (massless) photon
field $A_{\mu }$. After the superconducting phase transition, for
$T<T_{c}$, we are left with a real scalar field $\eta $ with mass
parameter $m_{\eta }^{2}=-2m^{2}$, accounting for the fluctuations
of the condensate, and a finite range electromagnetic field
described by a massive photon field $A_{\mu }$. The degree of
freedom represented by the field $\theta $ before the phase
transition has been absorbed in order to give a mass to the photon
\cite{higgs}.

For small fluctuations, the dynamical equations for the
electromagnetic field simplify to:
\begin{equation}
\bvec{\nabla}\cdot \bvec{E} = -m_{A}^{2}A^{0}, 
\quad
\frac{\partial \bvec{B}}{\partial t}+\bvec{\nabla}\times \bvec{B} = -m_{A}^{2} \bvec{A} , \label{19}
\end{equation}
and, after some passages, in the stationary
case we arrive at the following equations:
\begin{equation}
\nabla^{2}\bvec{E} = m_{A}^{2}\bvec{E},  \quad
\nabla^{2}\bvec{B} = m_{A}^{2}\bvec{B}.  \label{21}
\end{equation}
Let us focus on the equation for $\bvec{B}$ (the same applies to $\bvec{E}$) and
consider the simple one-dimensional case:
\begin{equation}
\frac{d^{2}B}{dx^{2}}=m_{A}^{2}B.  \label{22}
\end{equation}
The solution of this equation is:
\begin{equation}
B=B_{0}e^{-x/\delta }  \label{23}
\end{equation}
and we recover the well-known Meissner effect with a penetration
length \cite{tinkham}:
\begin{equation}
\delta =\frac{1}{m_{A}}\simeq \frac{\delta_{0}}{\sqrt{1-\frac{T}{T_{c}}}}.
\label{24}
\end{equation}
Instead, the dynamical equation for the fluctuation field $\eta $,
in the stationary, one dimensional case reduces to:
\begin{equation}
\frac{d^{2}\eta }{dx^{2}}=m_{\eta }^{2}\eta ,  \label{26}
\end{equation}
whose solution is again in the form:
\begin{equation}
\eta =\eta_{0}e^{-x/\xi^{\prime }}  \label{27}
\end{equation}
but with a coherence length
\begin{equation}
\xi^{\prime }=\frac{1}{m_{\eta }}=\frac{1}{\sqrt{-2m^{2}}}\simeq
\frac{\xi_{0}}{\sqrt{2(1-\frac{T}{T_{c}})}}. \label{28}
\end{equation}
In the lagrangian formalism outlined above we have thus recovered the basic
features of the supercoducting phase transition. Note also that the
fundamental parameters of the lagrangian theory, $\lambda $ and $\phi_{0}$
(besides the electric charge $e$), are directly related to the penetration
length $\delta $ of the magnetic field in the superconductor and to the coherence length of the condensate of the Cooper pairs
$\xi $ (or $\xi^{\prime }$), through Eqs. (\ref{24}) and (\ref{4c}) (or (\ref{28})).

In the GL theory the phase field $\theta $ has
disappeared from the description of the superconductor by means of
a gauge transformation on the field $A_{\mu }$. Such a
transformation does not affect the physics of the superconductor
as long as $A_{\mu }$ is massless, in which case the lagrangian
(and the dynamical equations) are gauge invariant. Instead, it is
well known that a massive $A_{\mu }$ field breaks out the gauge
invariance of the lagrangian (the mass term $m_{A}^{2}A_{\mu
}A^{\mu }$ is not gauge invariant), so that a gauge transformation
should alter, in principle, the dynamics of the system. In the
standard theory, the underlying physical motivation for
re-absorbing the field $\theta $ is that, for temperatures lower but not very far from the critical
one $T_{c}$, the phase fluctuations of the Cooper pairs are
irrelevant. Here we now relax this assumption and consider the
superconductor at very low temperatures, where it is expected that
phase fluctuations (that is, quantum fluctuations) become
relevant.

Coming back to the lagrangian in Eq. (\ref{11}) and restoring the
contribution of the $\theta $ field, we have to add to ${\cal L}_{GL}$ the
following quantities (neglecting third and fourth-order terms):
\begin{equation}
{\cal L}_{\theta }=\frac{1}{2}(\partial_{\mu }\theta )(\partial^{\mu
}\theta )+2\sqrt{2}e\phi_{0}A_{\mu }\partial^{\mu }\theta , \label{29}
\end{equation}
where the first term is the usual expression for the kinetic
energy of the (real) scalar field $\theta $, while the last one is
properly due to the Higgs mechanism and accounts for the
interaction of $\theta $ with the electromagnetic field. The EL
equation for the phase field is then:
\begin{equation}
\partial_{\mu }\partial^{\mu }\theta =-m_{A}\partial_{\mu }A^{\mu }.
\label{30}
\end{equation}
This equation is implicitly contained in the dynamical equations
for $A_{\mu }$, namely:
\begin{equation}
\partial_{\mu }F^{\mu \nu }+m_{A}^{2}A^{\nu }+m_{A}\partial^{\nu }\theta
=0.  \label{31}
\end{equation}

In fact, by taking the 4-divergence of (\ref{31}) and noting that
$\partial_{\nu }\partial_{\mu }F^{\mu \nu }=0$, we immediately re-obtain Eq. (\ref
{30}). Observe that, after the superconducting phase transition (i.e. for $%
m_{A}\neq 0$ and $\partial_{\mu }A^{\mu }\neq 0$), the evolution
of the phase field is not trivial since $\partial_{\mu
}\partial^{\mu }\theta \neq 0$, while the dynamics of $A^{\mu }$
(and, thus, the London equation) is not affected by the
interaction with the field $\theta$, as it is easily checked from
Eq. (\ref{31}). It, then, appears that the inclusion of the
$\theta$ degree of freedom does not alter the basic features of
superconductivity.

It is well known that an alternative description of the degrees of
freedom of a massless, real scalar field is provided by a
KR field \cite{kalb}, that is an antisymmetric tensor
$\theta_{\mu \nu }$ which generalizes to two indices the role of a
$U\left( 1\right) $ gauge potential $A^{\mu }$. The kinetic energy
of such a field is similar in form to that of the usual
electromagnetic field, namely $-\frac{1}{12}H_{\mu \nu \rho
}H^{\mu \nu \rho },$
where the field strength $H_{\mu \nu \rho }$ is obtained from the potential $%
\theta_{\mu \nu }$ by means of the totally antisymmetric (in the 3
indices) relation: $H_{\mu \nu \rho }=\partial_{\mu }\theta_{\nu
\rho }+\partial_{\nu }\theta_{\rho \mu }+\partial_{\rho
}\theta_{\mu \nu }$. It is simple to convince that, due to the
antisymmetry properties of $H_{\mu \nu \rho }$, the kinetic energy
identically equals the first
term in Eq. (\ref{29}), provided that the relation between the scalar field $%
\theta $ and the KR field is the following \cite{salvatore}:
\begin{equation}
\partial_{\mu }\theta =\frac{1}{6}\varepsilon_{\mu \nu \rho \sigma }H^{\mu
\nu \rho }=\frac{1}{2}\varepsilon_{\mu \nu \rho \sigma }\partial^{\nu
}\theta^{\rho \sigma }.  \label{34}
\end{equation}
As a consequence, the degrees of freedom of a massless, real scalar field $%
\theta $ are exactly the same of a massless, real KR
field $\theta_{\mu \nu }$ and, at least on the classical level,
the dynamics of non interacting fields can be alternatively
described by the two different approaches \cite {lund}.

Although the two descriptions give the same result (in the mentioned
conditions), a possible reason for considering a KR field in place of a
simple scalar field $\theta $ in the theory of superconductors is that an
antisymmetric tensor field seems more suitable for accounting vortices or
other topological effects in superconductors at very low temperatures. In
such a case, we have to replace the lagrangian in (\ref{29}) with the
following, making use of Eq. (\ref{34}):
\begin{equation}
{\cal L}_{KR}=-\frac{1}{12}H_{\mu \nu \rho }H^{\mu \nu \rho }+\frac{1}{%
6\delta }\varepsilon_{\mu \nu \rho \sigma }A^{\mu }H^{\nu \rho \sigma },
\label{35}
\end{equation}
where $\delta $ is the penetration length introduced above. The role of a KR field in the dynamics of superconductors and, in particular, a lagrangian similar to
that in Eq. (\ref{35}) has already been considered in the
literature \cite{balach}. However, the main
difference between these models and the present approach is that
the coupling of $H^{\mu \nu \rho }$ with the electromagnetic
potential $A^{\mu }$ is there
left completely arbitrary, while here is ruled by the penetration length $%
\delta $ and naturally emerges from the GL theory.

We now explore the consequences of replacing the lagrangian in Eq. (\ref{29}) with that in Eq. (\ref{35}). First of all we observe that, as long as the
non superconducting phase ($T>T_{c}$) is concerned, the two lagrangians give
the same result. In fact, in this phase the photon is massless ($m_{A}=0$),
so that from Eq. (\ref{30}) we obtain that the dynamics of $\theta $ is
given by $\partial_{\mu }\partial^{\mu }\theta =0$. This is consistent
with the identification (\ref{34}) since, by taking the 4-divergence of this
equation, its r.h.s. identically gives zero due to the antisymmetry of $%
\varepsilon_{\mu \nu \rho \sigma }$. The things change for
$T<T_{c}$, where the r.h.s. of Eq. (\ref{30}) does not vanish, so that
$\partial_{\mu }\partial^{\mu }\theta \neq 0$, which is now
inconsistent with the identification (\ref{34}). This takes place
because of the interaction term in (\ref{29}) or (\ref{35}): the
degrees of freedom of $\theta $ excited by the interaction with
$A^{\mu }$ are different from those of the KR field and,
as we will see below, this happens essentially because $H^{\mu \nu
\rho }$ develops additional (vector) degrees of freedom besides
those of a scalar field. We now assume that, contrary to what
happens in the standard GL
theory, the dynamics of the phase field is played by $H$ (rather than $%
\theta $) and described by ${\cal L}_{KR}$ in Eq. (\ref{35}) (rather than $%
{\cal L}_{\theta }$ in Eq. (\ref{29})). With this assumption we have not modified the description of the basics of superconductors, all of which are recovered in our model, but we have introduced
novel peculiar characteristics which we now study.

The addition of the lagrangian term (\ref{35}) into (\ref{11})
modifies the dynamical equations for the phase and the
electromagnetic field (Eqs. (\ref{30}) and (\ref{31})). After some
passages, the EL equations for $\theta_{\mu \nu } $ and $A_{\mu }$ look as follows:
\begin{eqnarray}
\partial_{\mu }H^{\mu \alpha \beta } &=&-\frac{1}{2\delta }\varepsilon^{\mu \nu
\alpha \beta }F_{\mu \nu }  \label{36} \\
\partial_\mu F^{\mu \nu }+\frac{1}{\delta^{2}}A^{\nu } &=&-\frac{1}{6\delta }%
\varepsilon^{\nu \mu \alpha \beta }H_{\mu \alpha \beta }. \label{37}
\end{eqnarray}
The different dependence on $\delta $ of the two terms in Eq.
(\ref{37}) makes the two corresponding contributions relevant at
different temperature regimes. In particular, for very small
$\delta $, the dominant contribution is that in
$\frac{1}{\delta^{2}}$, which drives the usual London equation for
superconductivity, namely Eq. (\ref{22}) or its solution
(\ref{23}). It, then, remains to investigate on the role of the
last term in (\ref{37}) and, at this end, \ we have to discuss
first the dynamics of the KR field $H^{\mu \nu \rho }$.
It is easy to see that, in the present scenario, there are 1
(scalar) +3 (vector) degrees of freedom carried by $H$, which can
be parameterized in terms of $J^{\nu
}=-\frac{1}{6}\varepsilon^{\nu \mu \alpha \beta }H_{\mu \alpha
\beta }$, or, writing $J^{\nu }=\left( \rho
,{\bvec{J}}\right) $,
\begin{equation}
\rho  =-H_{123}\,\,\\, {\bvec J} =\left(
H_{023},-H_{013},H_{012}\right) . \label{40}
\end{equation}
In fact, by expliciting the components of $H$, the equations (\ref{36})
become, after some algebra:
\begin{equation}
{\bvec \nabla }\times {\bvec J} =\frac{1}{\delta }
{\bvec B},  
\quad
\frac{\partial {\bvec J}}{\partial t}+{\bvec \nabla }\rho
=-\frac{1}{\delta }{\bvec E},  \label{42}
\end{equation}
where ${\bvec E},{\bvec B}$ are the electric and magnetic field
inside the superconductor \footnote{The quantity $J^\mu$ can be viewed as a KR 4-current, which is topologically conserved due to the antisymmetry of $H$.}.

Let us now come back to Eqs. (\ref{37}) and focus on the equation
for the magnetic field ${\bvec B}$ in the stationary case  in order to see the
modifications induced by the KR field in the London equation (similar considerations apply to the electric field). We
have:
\begin{equation}
\nabla^{2}{\bvec B}=\frac{1}{\delta^{2}}{\bvec B}-\frac{1%
}{\delta }{\bvec \nabla }\times {\bvec J}. \label{43}
\end{equation}
In the simple one-dimensional case, by introducing Eq. (\ref{42}), we get
the final result:
\begin{equation}
\frac{d^{2}B}{dx^{2}}=0,
\end{equation}
that is Eq. (\ref{23}) no longer holds and superconductivity
disappears. \\
In comparison with the standard GL theory, the
physical picture behind the model considered here, where the role
of the phase field is played by a KR field, is the
following. For temperatures below the critical one, but
not very far from it, only the scalar degrees of freedom contained
in $H$ (corresponding to $\theta $ in the GL theory)
are excited, so that ${\bvec J}$ does not contribute and the
system is a superconductor. For very low temperatures, however,
other (vector) degrees of freedom in $H$ (not present in the
GL theory) play into the game, and the net effect of
${\bvec J}$ is that of re-absorbing the photon mass contribution,
thus breaking the superconductivity. A reasonable motivation for
this scenario can be find in the known fact that an antisymmetric
tensor field is well suited for considering vortices \cite{balach}. Actually, it
is expected that, at very low temperatures, quantum vortices arise in condensed matter systems due to quantum phase effects. In the
present model, such effects are described by a KR field,
and the presence of vortices makes the superconductivity to
disappear. It is remarkable that what discussed above has simple and directly testable phenomenological consequences, so that future experiments aimed to study the behavior of superconductors at very low temperatures may easily falsify the model presented here.

We warmly thank G. Esposito, F. Lizzi and C. Nappi for useful discussions and suggestions.

\end{document}